\documentclass[preprint,showpacs,preprintnumbers,amsmath,amssymb]{revtex4-1}

\usepackage{epsf}
\usepackage{graphicx}
\usepackage{dcolumn}
\usepackage{bm}

\input{epsf}

\usepackage{bm}

\input{epsf}

\begin{document}

\preprint{Phys.Rev.B}

\title{Theory  of electron states in a twisted two-valley 2D system}

\author{ M.V. Entin$^{1}$ and  L.I. Magarill$^{1,2}$}
\affiliation{$^{1}$Institute of Semiconductor Physics, Siberian
Branch of the Russian Academy of Sciences, Novosibirsk, 630090, Russia
\\ $^{2}$ Novosibirsk State University, Novosibirsk, 630090, Russia}

\begin{abstract}
A system similar to gapped graphene (for example, fluorinated)  containing two or more electron valleys is considered. It is assumed that the  material has a sector cut and is deformed in the plane and the the cut edges are connected to form an adiabatically curved atomic net without extended defects. We neglect the deformation potential. In such a system, the local momentum of the valley center ${\bf K}$ acts as the vector potential of fictitious magnetic field. We found the electron states  in such  system in the case of orientation ${\bf K}$ along the azimuth of geometric space  at any point. It is shown that the vector potential results in the appearance of local discrete   electron states.
Mathematically, the problem is mapped onto the Coulomb problem with an effective charge depending on ${\bf K}$.
\end{abstract}

\maketitle
\section*{Introduction}
The purpose of the present paper is studying of electron states in an adiabatically and statically deformed 2D crystal. There are different
manifestations of such  deformation. One can mention the appearance of the geometric potential \cite{da costa}, \cite{ogawa} and  the modifications of the 2D Hamiltonian caused by curvilinearity \cite{chap&mag},\cite{ent&mag1},\cite{ent&mag2}. However, systems considered in above mentioned articles   possess simple energy spectrum.  Here  we concentrate on  multivalley materials.

 There are different kinds of deformation. In an open system the deformation is supported by an external field. The other kind of deformation, which is the topic of the present paper, is the situation when a stressed state is caused by a contact splice  of cut  edges. For the total
adiabaticity of the system the abutment joint should  conserve the crystallographic order. We assume that the deformed material  is, locally, a crystal and  has no extension defects anywhere.  This condition can be formulated as if one goes around a cut vertex in a closed loop in real space, and  the material should coincide with itself (see Fig.1). This means that after such  walk, the material changes in accordance with the crystallographic group.
\begin{figure}[ht]
\centerline{\epsfysize=5.cm\epsfbox{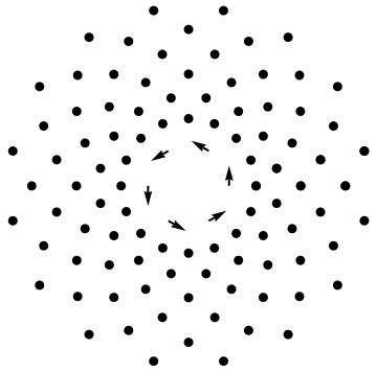}}
\caption{Adiabatically deformed gapped graphene. Vector ${\bf K}({\bf r})$  (arrows) experiences a revolution with the local crystal axes, being directed along the azimuth.} \label{Fig1}
\end{figure}

\section*{Problem formulation}
The number of valleys is determined by the crystal symmetry. In particular, fluorinated graphene or dichalcogenide have two independent valleys which can be numerated by the valley centers ${\bf K}$ and $-{\bf K}$  in the Brilllouin zone.  To specify the valley further, we will use valley index $\tau=\pm 1.$ These ideal systems in the  single-band (electron states in the conduction band will be studied) envelope approximation are described by the Hamiltonian
\begin{equation}\label{Hamilt}
 H=({\bf\hat{ p}-\tau K})^2/2m_e,
\end{equation}
where $\hat{\bf p}$ is the 2D momentum operator, $m_e$ is the effective mass in the conduction band; we set $\hbar=1$.

We consider the situation with no electric or magnetic fields. Now let us make ${\bf K}$ to depend on planar coordinates $(x,y)={\bf r}$, ${\bf K}={\bf K}({\bf r})$.

In the deformed system the Hamiltonian Eq.(\ref{Hamilt}) changes as
\begin{equation}\label{Hamilt_def}
 H=({\bf\hat{ p}-\tau K}({\bf r}))^2/2m_e.
\end{equation}
Assume that ${\bf K}({\bf r})\cdot {\bf r}=0$ and ${\bf K}({\bf r})\parallel [{\bf n}_z,{\bf r}]$. Here   ${\bf n}_z$ is the unit vector of the normal to the system plane.

In  polar coordinates $(r,\varphi)$ the Schr\"odinger equation can be written  as
\begin{equation}\label{Scheq_pol}
-\frac{1}{r}(\partial_r(r\partial_r\Psi)-(\frac{1}{r}\partial_\varphi-i\tau K)^2\Psi=2m_eE\Psi,
\end{equation}
where ~~ $K=|{\bf K}|=const$; energy $E$ is counted from the conduction band bottom.

\section*{Coulomb-like states}
Let us present the wave-function as $\Psi(r,\varphi)= R(r) \exp{(im\varphi )}/\sqrt{2 \pi}$, where $m=0, \pm 1,\pm 2 ...$. Then angle variable $\varphi$ is separated and we arrive at the equation for radial function $R(r)$:
\begin{equation}\label{Scheq_rad}
\frac{d^2R}{dr^2} + \frac{1}{r}\frac{dR(r)}{dr} +\Bigl\{2m_e\bigl[E-\frac{1}{2m_e}(K^2+\frac{m^2}{r^2})\bigr] +\frac{m \tau K}{r}\Bigr\}R(r)=0.
\end{equation}
One can see that Eq.(\ref{Scheq_rad}), in fact, is the Schr\"odinger equation for the 2D Coulomb center with a potential of the form $-\alpha/r$, in which  value $-m \tau K/me$ acts as  coefficient  $\alpha$. We can immediately conclude that for $m \tau < 0$, i.e. when the potential is attractive, localized states appear.

Note that the states appear in pairs with $\tau =+1$ and $\tau =-1$. Hence, localized Coulomb-like states with positive $m$ appear if $\tau=-1$  and {\it vice versa}. For the state $m=0$ the effective charge vanishes and localized states disappear.
The solution of 2D Coulomb problem was found,  in  \cite{chap&ent},\cite{yang}.
Using the results of  these publications we have the following expressions for the energies of discrete states
\begin{equation}\label{spectr}
E_{n,m}=\frac{K^2}{2m_e}\left(1-\frac{m^2}{(n -1/2)^2}\right),~~~~~ |m|=0,1,2,...n-1,
\end{equation}
where $n=1,2,3...$ is the principal quantum number.
 The normalized radial functions are given by
 \begin{eqnarray}\label{rad_func}
   &R_{n,m}^{(\tau)} = \frac{\beta_{n,m\tau}}{(2|m|)!}\Bigl[\frac{(n+|m|-1)!}{(2n-1)((n-|m|-1)!)}\Bigr]^{1/2} (\beta_{n,m\tau} r)^{|m|}\exp{(-\beta_{n,m\tau} r/2)} \times \nonumber \\& _1F_1(-n+|m|+1,~ 2|m|+1,~ \beta_{n,m\tau} r); ~~~~~\beta_{n,m\tau}=\frac{2m_e |m\tau |K}{n-1/2};~~~~~~~~(m\tau < 0).
   \end{eqnarray}

\begin{figure}[ht]
\leavevmode\centering{\epsfxsize=7cm\epsfbox{ 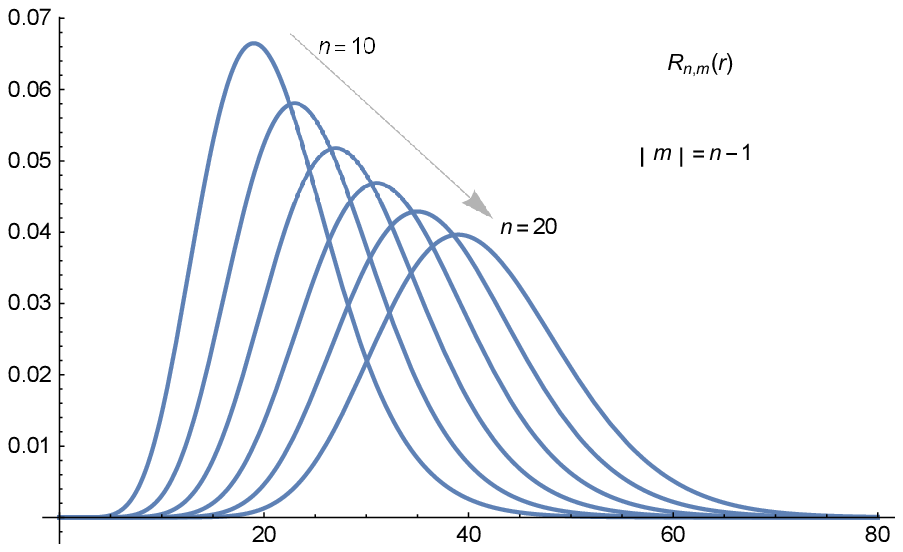}}
\leavevmode\centering{\epsfxsize=7cm\epsfbox{ 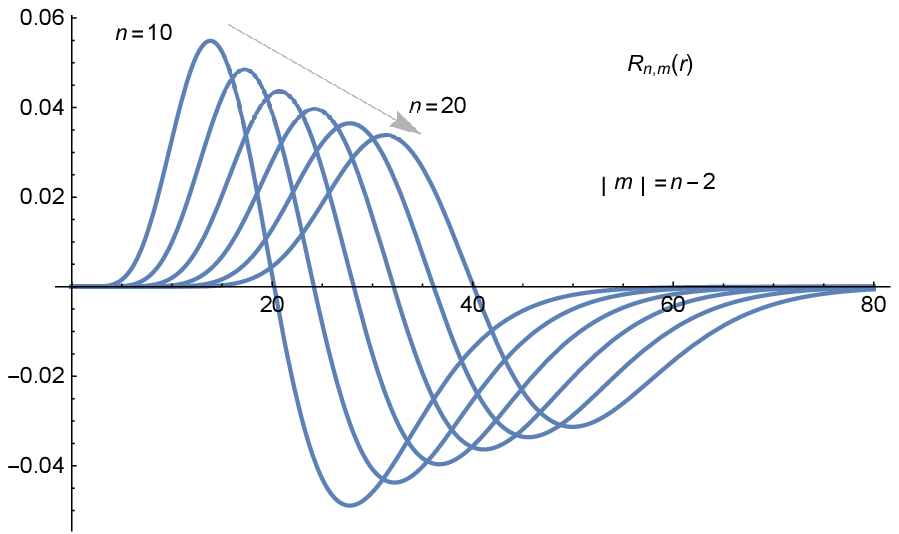}}
\leavevmode\centering{\epsfxsize=7cm\epsfbox{ 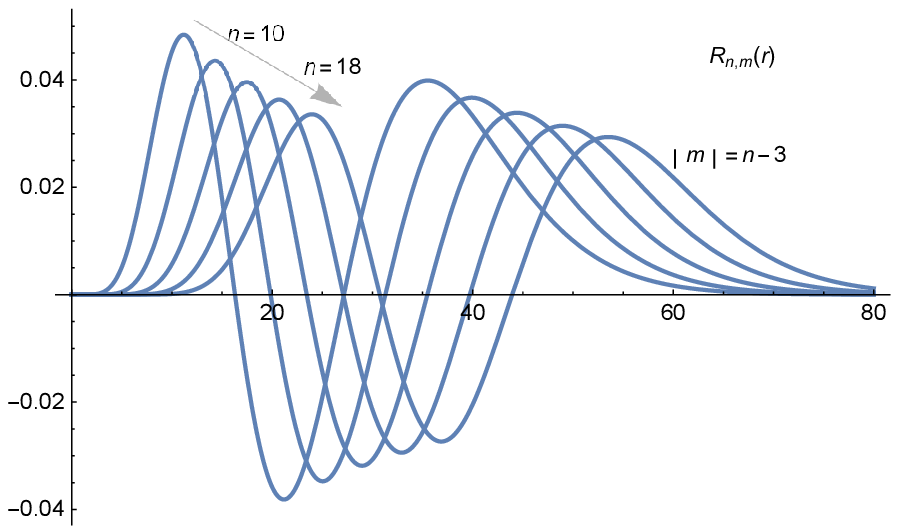}}
\caption{ Radial wavefunctions at different $m$ and $n$.}\label{Fig2} \end{figure}

 The characteristic order of the $E_{n,m}$ magnitude
is $K^2/(2m_e)$.
This is a very large quantity which is  beyond the applicability domain of the simple model of Eq.(\ref{Hamilt}). However, if $|m|\approx n-1/2 $ and $|m| \gg 1$, then    $E_{n,m}\to 0$, and that revives the envelope-function assumptions.
If so,
 \begin{equation} \label{spectr_appr} E_{n,m}\approx \frac{K^2}{m_e}\frac{n-1/2-|m|}{n-1/2}.\end{equation}
 Note, that although the states $|m|$ and $-|m|$  are degenerate, they belong to different valleys.
 These states have positive energies, but they are discrete and  localized. This is an unusual, but not forbidden situation. In fact, the motion in the coordinate space automatically changes the relation of electron momentum and the local valley minimum; that blocks electron relocation and localizes the electron states.

 Another remark concerns the spectrum discreetness. As far as  energies are proportional to the rational expression from numbers $n$ and $|m|$, the spectrum covers all positive energies $\ll K^2/2m_e$.

Thus, essentially different electron spectrum shows the important influence of the valley center coordinate dependence ${\bf K}({\bf r})$ on the electron states.

 \section*{Optical transitions}

Let us illuminate the system by some normally directed circular-polarized light  which is resonant  with respect to some interlevel distance.
     To determine the interlevel transition probability
it is necessary to calculate the matrix elements of
 operator  $({\bf \hat{v}e})$, where ${\bf \hat{v}} = ({\bf \hat{p}}-\tau K({\bf r}))/m_e$ is the velocity
operator corresponding to the Hamiltonian Eq.(\ref{Hamilt}), and ${\bf e}$
is the electromagnetic wave polarization vector.

The matrix elements are expressed via the integrals
of the radial functions and, for circular polarization ${\bf e}=(1,~i\zeta)/\sqrt{2} ~~(\zeta=\pm 1)$, they are given
 \begin{eqnarray}\label{matelem}
 \langle \Psi_{n',m'}^{(\tau)*}|({\bf \hat{v}e})|\Psi_{n,m}^{(\tau)}\rangle = \delta_{m',m+\zeta} Z_{n',n;m}; \nonumber \\
  Z_{n',n;m}^{(\tau)} = \int_0^\infty dr r R_{n',m+\zeta}^{(\tau)}(r)\Bigl[ -i\frac{R_{n,m}^{(\tau)}(r)}{dr} +\bigl(\frac{im\zeta}{r}-i\tau\zeta K\bigr)R_{n,m}^{(\tau)}(r)\Bigr].
\end{eqnarray}
Eq.(\ref{matelem}), together with the wave functions Eq.(\ref{rad_func}), allows easy to calculate the oscillator strengths for different optical transitions.

Taking into account that the light causes dipole transitions, we have $E_{n,m}-E_{n',m+1}=\omega$ for, say,  $\zeta=+1$.  If $m>0$. the transitions occur between localized states in the valley $\tau=-1$. At the same time another valley will stay inactive. In other words, in dependence on the polarization sign (i.e., on sign of $\zeta$), the selected valley will  be active or not with respect to transitions between discrete levels.

Thus, we can selectively excite electrons in the  valleys ${\bf K}$ or $-{\bf K}$. This property unifies the considered system with non-deformed dichalcogenides. However, in the deformed crystal the excitation touches the localized states, instead of 2D extended states in non-deformed dichalcogenides.

Note, that the transitions in non-active valley occur also, but they have no resonant nature; this means much lower absorption at  frequency corresponding to discrete states.

\section*{Generalizations}
\subsection*{The case ${\bf K}({\bf r})||{\bf r}$.}
The obtained results can be simply generalized. First, one can consider another configuration of  vector ${\bf K}$, when  ${\bf K}({\bf r})||{\bf r}$. In this case ${\bf K(r)}$ represents a potential field, ${\bf K(r)}= \nabla U(\bf r)$, and can be excluded from the wave function by the gauge transformation. Thus, ${\bf K}({\bf r})||{\bf r}$ does not influence
 electron states.

\subsection*{Two-bands narrow-gap model.}
Another generalization concerns the model. In fact, the considered basic materials have relatively close conduction (c) and valence (v) energy bands. The previous consideration is valid for an electron state close to the conduction band bottom. To extend the model  one should base  on the minimal two-band model \cite{di xiao, falko}. Taking into account the dependence of  vector  ${\bf K({\bf r})}$ on ${\bf r}$,  the Hamiltonian of this model becomes
\begin{eqnarray}\label{Ham_min model}
\hat{H}=&\left(
    \begin{array}{cc}
      0 & \gamma[(\hat{p}_x-\tau K_x)-i(\hat{p}_y-\tau K_y)]\\
     \gamma[(\hat{p}_x-\tau
     K_x)+i(\hat{p_y}-\tau K_y)] & -\Delta \\
    \end{array}
  \right).
\end{eqnarray}
Here $\gamma$ is the interband velocity,  $\Delta$ is the gap between c- and v-bands.
The eigenfunction of Eq.(\ref{Ham_min model}) is the two-component spinor $\Psi({\bf r})=(\Psi_1({\bf r}),\Psi_2({\bf r})).$
Eliminating $\Psi_2$ from the system of equations $\hat{H}\Psi = E\Psi$, we find:
\begin{eqnarray}\label{eq_min}
\Bigl\{\frac{({\bf p}-{\bf K({\bf r})})^2}{2m_e} -\frac{E(\Delta  + E)}{\Delta}-\frac{\tau}{2m_e}({\bf n}_z\cdot[{\bf \nabla}\times{\bf K({\bf r})}])\Bigr\} \Psi_1({\bf r})=0.
\end{eqnarray} We have used relation $m_e=\Delta/2\gamma^2$.
For the case, ${\bf K({\bf r})}=[{\bf n}_z\times {\bf r}]K/r$, Eq.(\ref{eq_min}), after separating the angular variable ($\Psi_1({\bf r})=R_1(r) \exp{(im\varphi)}$), is  transformed into
\begin{equation}\label{eq_min1}
\frac{d^2R}{dr^2} + \frac{1}{r}\frac{dR(r)}{dr} +\Bigl\{2m_e\bigl[\frac{E(\Delta  + E)}{\Delta}-\frac{1}{2m_e}(K^2+\frac{m^2}{r^2})\bigr] +\frac{(m+1) \tau K}{r}\Bigr\}R(r)=0; ((m+1)\tau <0).
\end{equation}
Comparing Eq.(\ref{eq_min1})  with Eq.(\ref{Scheq_rad})  we  obtain the expression for the spectrum in the two-band case:
\begin{eqnarray} \label{spectr_minmax}
  E_{n,m}^{(c,v)} &=& -\frac{\Delta}{2} \pm\sqrt{\frac{\Delta^2}{4}+\frac{K^2}{2m_e}\Delta (1-\frac{(m+1)^2}{(n-1/2)^2})},
  \end{eqnarray}
or expanding with respect to large $ n$ and $m$:
\begin{eqnarray} \label{spectr_approx}E_{n,m}^{(v)} &=& -\frac{\Delta}{2} -\sqrt{\frac{\Delta^2}{4}+\frac{K^2}{m_e}\Delta \frac{n-m-3/2}{n-1/2}}. \end{eqnarray}
\subsection*{Gapped graphene cone}
Conic graphene ripples were realized in \cite{bouchiat-bouchiat1} and \cite{bouchiat-bouchiat2}. From the theoretical point of view,  a disadvantage of planar deformed gapped graphene is a strong deformation of external graphene parts. Conic graphene is a system where the in-plane deformation vanishes. Such system is similar to a inextensible sheet of paper rolled into a cone. In fact, this system is equivalent to planar graphene. Here we consider the conic gapped graphene within the single-band approximation.

To fulfill the adiabaticity  requirement one can cut  sector $2\pi/3$ from a graphene disk with radius $r_0$, then connect the edges of the cut. As a result, we obtain a cone where the crystal structure is continuously repeated through the cut (see  Fig.(\ref{cone}). The cone height  is $\sqrt{8}r_0/3$, the base radius  is $r_0/3$.
\begin{figure}[ht]\label{cone}
\leavevmode\centering{\epsfxsize=7cm\epsfbox{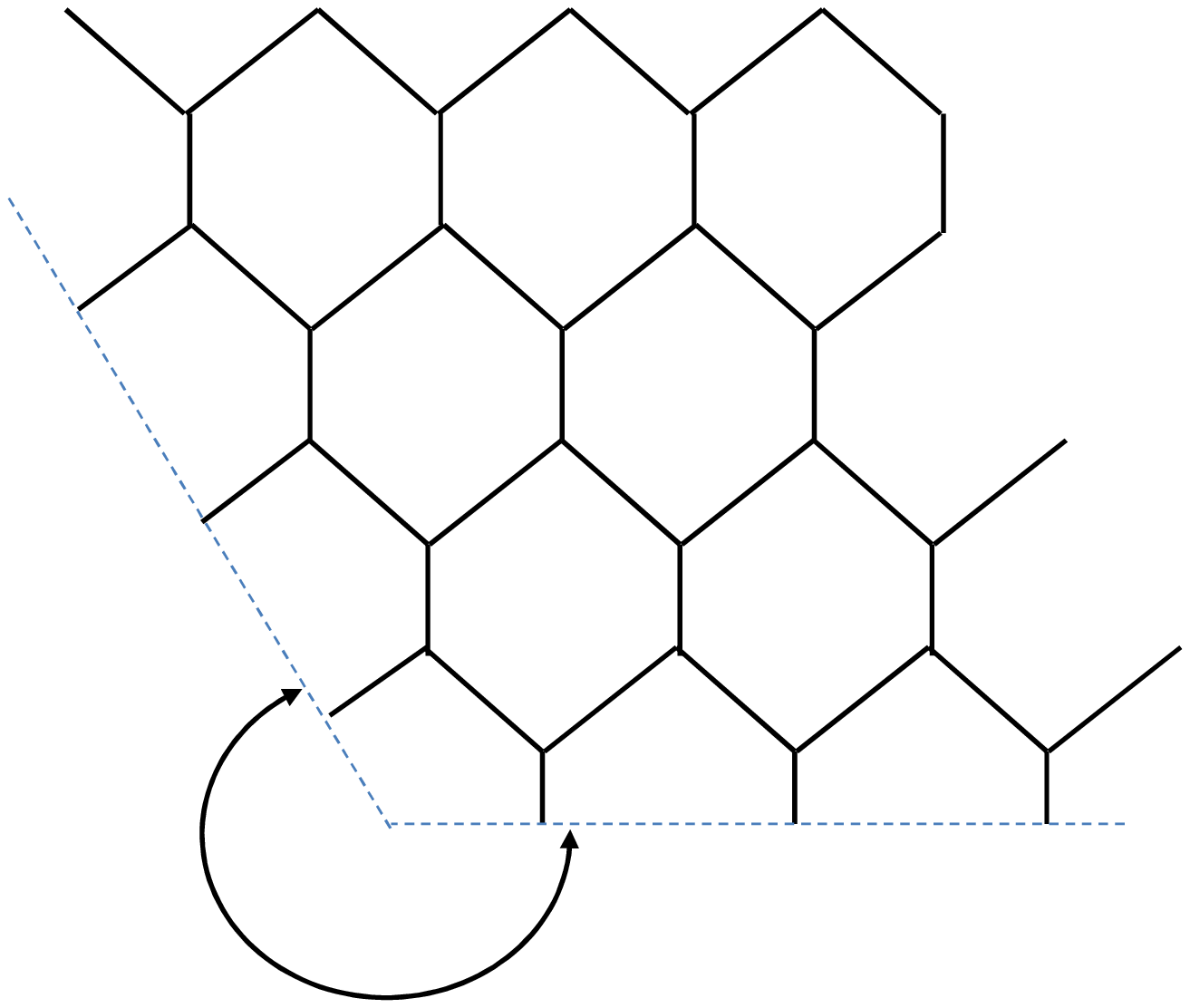}}
\leavevmode\centering{\epsfxsize=7cm\epsfbox{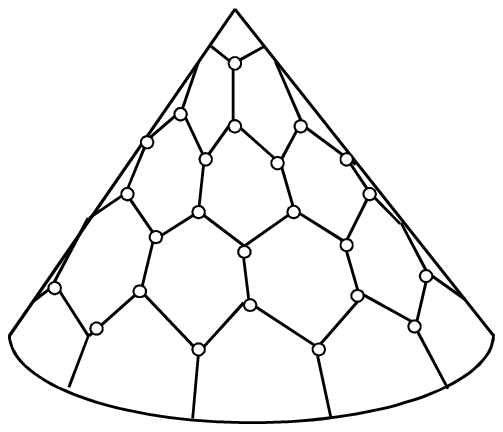}}
\caption{The graphene cone (right panel) is formed from  a list (left panel) by a connection of rays $y=0, ~x>0$ and $y=-x\sqrt{3}/2,~x<0$, what guarantee the absence of extended defects. }\label{Fig2} \end{figure}

Mathematically, in the envelope approximation, electrons in gapped graphene on the
surface can be described  by the Schr\"odinger equation in spherical coordinates
\begin{equation}\label{cone}
\frac{1}{\rho^2}\frac{\partial}{\partial \rho}\left(\rho^2\frac{\partial }{\partial \rho}\right)\Psi+ \left(\frac{1}{ \rho\sin\theta}\frac{\partial}{\partial\phi}-i\tau K\right)^2\Psi+2m_eE\Psi=0.
\end{equation}
Here $\theta=const$ is the half-angle at the cone apex, $\rho$ is the distance to the apex. In the cone shown in Fig.(\ref{cone}) $\sin(\theta)=1/3$. After the separation of angle $\phi$ we come to

\begin{equation}\label{Scheq_rad2}
\frac{d^2R}{d\rho^2} + \frac{1}{\rho}\frac{dR(\rho)}{d\rho} +\Bigl\{2m_eE-K^2-\frac{\tilde{m}^2-1/4}{\rho^2} +\frac{\tilde{m} \tau K}{\rho}\Bigr\}R(\rho)=0,
\end{equation}
where $\tilde{m}=m/\sin(\theta)$. The replacement $m^2\to \tilde{m}^2-1/4 $ and $\tilde{m}K\to mK$  converts  Eq.(\ref{Scheq_rad2}) into Eq.(\ref{Scheq_rad}).

\section*{Summary}
In conclusion,  we have found the electron states in a planar deformed multivalley 2D system.
Unexpectedly, the influence of the adiabatic deformation turned to be strong. One can consider this new interaction as a topological one. The characteristic energy of this interaction is based on  quantity $K^2/m_e$. This quantity has the value  of the order of several $eV$, and it exceeds the deformation potential, which reaches such  value at an extreme linear deformation $\sim 0.1$ only. Also, this topological interaction exceeds the characteristic geometric potential (which entirely vanishes for planar deformation,  but can be nonzero  for the space deformation of a  plane of the system).  Obviously, there is a lot of different other deformation kinds of 2D systems. However, our finding shows the important role  of the  vector ${\bf K({\bf r})}$ rotation, considered in the present paper.

\section*{Acknowledgements} This work has been supported  by RFBR grant No 20-02-00622.

\end{document}